\documentclass[12pt, draftclsnofoot, onecolumn]{IEEEtran}
\usepackage{amsmath,amssymb,amsfonts,bm}
\usepackage{algorithmic}
\usepackage{algorithm}
\usepackage{array}
\usepackage[caption=false,font=normalsize,labelfont=sf,textfont=sf]{subfig}
\usepackage{textcomp}
\usepackage{stfloats}
\usepackage{bigints}
\usepackage{url}
\usepackage{verbatim}
\usepackage{graphicx}
\usepackage{mathtools}
\usepackage{dsfont}
\usepackage{esvect}
\usepackage{accents}
\usepackage{cite}
\usepackage{xcolor}
\newcolumntype{M}[1]{>{\centering\arraybackslash}m{#1}}
\newcommand*{\QEDA}{\null\nobreak\hfill\ensuremath{\blacksquare}}%

\newcommand\myeqa{\mathrel{\stackrel{\makebox[0pt]{\mbox{\normalfont\scriptsize (a)}}}{=}}}
\newcommand\myeqb{\mathrel{\stackrel{\makebox[0pt]{\mbox{\normalfont\scriptsize (b)}}}{=}}}
\newcommand\myeqc{\mathrel{\stackrel{\makebox[0pt]{\mbox{\normalfont\scriptsize (c)}}}{=}}}
\newcommand\myeqd{\mathrel{\stackrel{\makebox[0pt]{\mbox{\normalfont\scriptsize (d)}}}{=}}}

\newcommand\myeqe{\mathrel{\stackrel{\makebox[0pt]{\mbox{\normalfont\scriptsize (e)}}}{=}}}

\newcommand\approxequiv{\mathrel{\stackrel{\makebox[0pt]{\mbox{\normalfont\scriptsize approx.}}}{\sim}}}
\begin{document}

% paper title
% Titles are generally capitalized except for words such as a, an, and, as,
% at, but, by, for, in, nor, of, on, or, the, to and up, which are usually
% not capitalized unless they are the first or last word of the title.
% Linebreaks \\ can be used within to get better formatting as desired.
% Do not put math or special symbols in the title.
\title{Joint Energy and SINR Coverage Probability in UAV Corridor-assisted RF-powered IoT Networks}
%  
%
% author names and IEEE memberships
% note positions of commas and nonbreaking spaces ( ~ ) LaTeX will not   brea k
% a structure at a ~ so this keeps an author's name from being broken across
% two lines.
% use \thanks{} to gain access to the first footnote area
% a separate \thanks must be used for each paragraph as LaTeX2e's \thanks
% was not built to handle multiple paragraphs
%

\author{Harris K. Armeniakos,~\IEEEmembership{Graduate Student Member,~IEEE,}\\
        Petros S.~Bithas,~\IEEEmembership{Senior Member,~IEEE,}\\
        Konstantinos~Maliatsos,~\IEEEmembership{Member,~IEEE,}
       and \\ Athanasios G.~Kanatas,~\IEEEmembership{Senior Member,~IEEE}
        % <-this % stops a space
\thanks{H. K. Armeniakos and A. G. Kanatas are with the University of Piraeus, Piraeus, Greece (e-mail: \{harmen,kanatas\}@unipi.gr). P. S. Bithas is with the National and Kapodistrian University of Athens, Athens, Greece (e-mail:pbithas@dind.uoa.gr). K. Maliatsos is with the University of the Aegean, Samos, Greece (e-mail: kmaliat@aegean.gr).}

% <-this % stops a space
}

\maketitle

% As a general rule, do not put math, special symbols or citations
% in the abstract or keywords.
\begin{abstract}
This letter studies the joint energy and signal-to-interference-plus-noise (SINR)-based coverage probability in Unmanned Aerial Vehicle (UAV)-assisted radio frequency (RF)-powered Internet of Things (IoT) networks. The UAVs are spatially distributed in an aerial corridor that is modeled as a one-dimensional (1D) binomial point process (BPP).  By accurately capturing the line-of-sight (LoS) probability of a UAV through large-scale fading i) an exact form expression for the energy coverage probability is derived, and ii) a tight approximation for the overall coverage performance is obtained. Among several key findings, numerical results reveal the optimal number of deployed UAV-BSs that maximizes the joint coverage probability, as well as the optimal length of the UAV corridors when designing such UAV-assisted IoT networks. 
\end{abstract}

% Note that keywords are not normally used for peerreview papers.

% For peer review papers, you can put extra information on the cover
% page as needed:
% \ifCLASSOPTIONpeerreview
% \begin{center} \bfseries EDICS Category: 3-BBND \end{center}
% \fi
%
% For peerreview papers, this IEEEtran command inserts a page break and
% creates the second title. It will be ignored for other modes.
\IEEEpeerreviewmaketitle

\section{Introduction}
As we step toward the sixth-generation (6G) communication networks, the UAVs are envisioned to constitute a core pillar of the IoT networks aiming to realize massive connections among several devices \cite{IoT}. In various use cases, the UAVs are considered to cooperatively operating in swarms to serve IoT devices either for communication purposes or for energy transfer \cite{swarms}. To this aim, in order to assure the safe and legal operation of the multiple drones, the concept of virtual air corridors with flight paths has been emerged by the Federal Aviation Administration (FAA) and the National Aeronautics and Space Administration (NASA) \cite{corridors}. These \textit{UAV corridors} are specific sky routes for UAVs to be distributed between the endpoints.  However, in these corridors, congestion of UAV swarms is expected, whose consequences can be alleviated by using lanes and rules for the autonomous mobility and movement of the UAVs, similar to the highway and traffic code regulating the coexistence of vehicles on the roads \cite{Access}. 

Over the last years, stochastic geometry has emerged as a very powerful tool for modeling and analyzing the performance of complicated aerial networks leading to tractable results and revealing valuable system-level insights. Nevertheless, few works  have employed stochastic geometry tools to investigate the performance achieved in UAV networks in terms of joint energy and communication coverage probability. In \cite{UAVharvesting1}, the authors studied the energy harvesting and the coverage probability analysis in a two-hop multiple UAV network with backhauling capabilities at terrestrial Base Stations (BSs). In \cite{UAVharvesting2}, the authors  investigated the joint probability of energy and SNR coverage in UAV laser-powered networks. However, most works studying the energy harvesting, e.g., \cite{UAVharvesting1}, \cite{Kishk}, \cite{Dhillon}, are based on an approximation of the mean value of the received power from all nodes except the dominant one, an approach that offers tractability with reduced, however, accuracy. Nevertheless, the introduction of the UAV corridors framework in UAV swarms scenarios, in which communication and/or energy transfer to IoT devices is expected, is a totally new research idea that leads to a \textit{UAV corridor-assisted RF-powered IoT network}.

\emph{Contributions:} The introduction of UAV swarms in RF-powered IoT networks requires a more nuanced spatial modeling of UAVs to sophisticatedly enable safe, efficient, and coordinated use of airspace by UAVs while minimizing the risk of collisions. In this letter, this research gap is addressed by the introduction of \emph{UAV corridors}, where the UAVs can be modeled as a \emph{1D BPP}. Notably, the 1D BPP is a novel point process for modeling the spatial locations of UAV-BSs in aerial corridors. Subsequently, the joint energy and SINR-based coverage probability is analytically derived under the presence of \emph{shadowing}. Note that unlike conventional approaches, which are based on a simplified approximation of the received power mean value, in this work the Gamma distribution approximation is exploited. As it will be shown, the proposed approach yields to an extremely tight approximation of the exact expression for the energy coverage probability. The results reveal several system-level insights, such as the optimal number of deployed UAV-BSs that maximizes the joint coverage probability, as well as the optimal length of UAV corridors, which has not been investigated in the existing literature.

\section{ System Model }
\subsection{Network Model}
Consider a downlink finite UAV-assisted IoT network that integrates both terrestrial IoT receiving devices and UAV-BSs. The IoT devices are assumed to be uniformly and independently distributed according to a stationary point process in a finite area $\mathcal{A} \subset \mathbb{R}^2$. For this setup, it is assumed that $\mathcal{A}$ is a two-dimensional (2D) circular area $b(\mathbf{o},R)$ centered at the origin $\mathbf{o} = [0,0]^{T}$ with radius $R$, i.e. $\mathcal{A}=b(\mathbf{o},R)$ as shown in Fig. 1. Next, consider $N$ UAV-BSs hovering in a UAV \textit{aerial corridor}, with rules enforced by appropriate traffic authorities. The UAV corridor is modeled as a line segment of length $2 R$, located $h$ meters above the ground. Without loss of generality, the ground projection of the UAV corridor crosses the origin of the coordinates $\mathbf{o}$, namely $\mathbf{o^{'}}=[0,0,h]^{T}$. Accordingly, the $N$ UAV-BSs are assumed to be uniformly and independently distributed on the line segment of length $2 R$ and their spatial locations form a 1D BPP $\Psi$. It is noted that such a system model has never been investigated in the past due to the absence of the UAV-corridor stochastic geometry framework. Subsequently, let $\{\mathbf{y}_i\} \equiv \Psi$ denote the spatial locations of UAV-BSs. Without loss of generality, it is  assumed that the receiving IoT device is located at the ground origin $\mathbf{o}$ at a given time. After averaging the performance of this receiving IoT device over $\Psi$, it becomes the typical IoT receiver, or simply \textit{receiver}. Next, let $\{d_i\}$ denote the unordered set of Euclidean distances between the receiver and the UAV-BSs. Then, the ordered set of Euclidean distances in ascending order is denoted by  $\{r_n\}_{n=1:N}$, where $r_n$ denotes the distance between the receiver and the $n$-th nearest UAB-BS. A representative illustration of the aforementioned model is shown in Fig. 1.

\subsection{Path Loss and Channel Models}

\subsubsection{Path loss}
The  channel conditions are characterized by different path-loss exponents, denoted by $\alpha$. Then, following the standard power-law path-loss model for the path between the receiver and a UAV-BS located at $\mathbf{y}_i \in \Psi$, the random path loss function  is defined as
\begin{equation}
l(\|\mathbf{y}_i\|) = K \|\mathbf{y}_i\|^{-\alpha} ,  
\end{equation}
where $K = \Big(\frac{c}{4 \pi f_c}\Big)^2$, with $c$  being the speed of light and $f_c$ the carrier frequency.

\subsubsection{Small-scale fading}
The channels amplitudes are assumed to experience Nakagami-$m$ fading  with fading parameter $m$. Therefore, the channels power gains $h_{i}^j$ follow Gamma distribution, while $j \in \{c,h\}$ is used to characterize the channel during the communication and energy harvesting phase, which are assumed to be independent. Therefore, $h_i^j \sim $Gamma$\big(m,\frac{1}{m}\big)$ with the shape and scale parameters of $h_i^j$ being $m$ and $1/m$, respectively. Note that $\mathbb{E}[h_i^j]=m \frac{1}{m}=1$. The probability density function (PDF) of $h_i^j$ is given by
\begin{equation}
  f_{h_i^j}(x) = \frac{{m}^{m} x^{m-1}}{\Gamma(m)}\exp{(-m x)},
\end{equation}
where  $\Gamma(\cdot)$ is the Gamma function \cite[eq. (8.310.1)]{Ryzhik}.

\subsubsection{Large-scale fading}
In several cases, due to the presence of obstacles between a UAV-BS and the receiver, e.g., buildings, trees, moving cars, the received signal is also subjected to shadowing. These shadowing random fluctuations are modeled by the inverse-gamma (IG) distribution, which has been recently proposed in UAV-to-ground communication scenarios \cite{Bithas}. Based on this approach, the non-line-of-sight scenarios due to random blockage are also accounted. The PDF of the IG distribution is given by
 \begin{equation}
f_{S_i^j}(x) =  \frac{\gamma^q}{\Gamma(q) x^{q+1}} \exp\Big(-\frac{\gamma}{x}\Big), 
\end{equation}
for $i=1,...,N$, $j \in \{ c,h\}$ and  $q > 1$ is the shaping parameter of the distribution, related to the severity of the shadowing, i.e., increased values of $q$ result in higher NLoS probability, whereas $\gamma$ denotes the scaling parameter. It is assumed that $S_i^c, S_i^h$ during the harvesting and communication phases, share the same shadowing parameters but they are independent random variables (RVs).

%\begin{figure*}
%  \includegraphics[width=.6\textwidth]{system_model_harvesting.eps}
%\centering
%  \caption{Illustration of the time-slotted architecture of the UAV-corridor assisted RF-powered IoT network. }
%\end{figure*}

\begin{figure}[!t]
    \centering
    \includegraphics[keepaspectratio,width=4in]{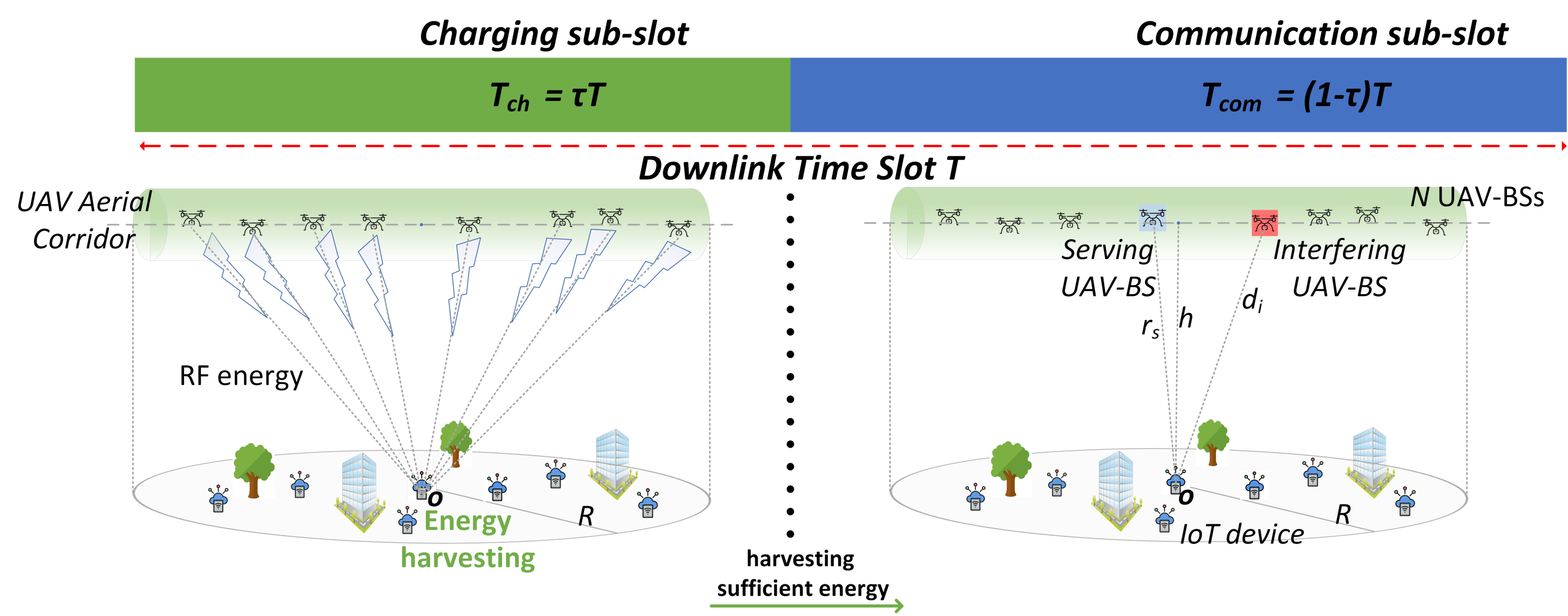}
       \caption{ Illustration of the time-slotted architecture of the UAV corridor-assisted RF-powered IoT network. }
    \label{fig:Fig2}
\end{figure}

\subsection{Time-Slotted Architecture}
A time-slotted architecture is adopted for the channel sharing between charging and communication functionalities. It is assumed that the RF signals is the only energy source of IoT devices. Each time slot of duration $T$ is divided into a charging sub-slot
with $T_{ch} = \tau T$ and a communication sub-slot with $T_{com} = (1-\tau) T$, where $\tau$ denotes the fraction of the downlink time slot devoted to energy harvesting. Next, the operations in each sub-slot are presented. 

%However, not all UAV-BSs are active at the considered time slot duration. Let $t_i$ denote the state of the $i$-th interfering UAV-BS that is, active denoted by "1", or inactive denoted by "0". In this work, $t_i$ is modeled as a discrete random variable  with probability mass function (pmf) given by 
%\begin{equation}
%p_{t_i} = \left\{\begin{array}{ll} 
%  1,  & \mbox{w.p.} \, \,  p_{on} \\
% 0,  & \mbox{w.p.} \, \, p_{of\hspace{-1.2 pt}f} = 1-p_{on}, 
%\end{array}\right.
%\end{equation}
%for $i = 1,...,N$, where $p_{on}$ and $p_{of\hspace{-1.2 pt}f}$ denote the probability that the $i$-th interfering UAV-BS be active and inactive, respectively.

\subsubsection{Charging sub-slot}
In the charging sub-slot, all active UAV-BSs act as RF energy sources for IoT devices. The energy harvested in this sub-slot, $E_h$, should be greater than the minimum energy demand $\gamma_h$ to make the transmission succeed, i.e., $E_h \geq \gamma_h$, where $E_h$ can be expressed as
\begin{equation}
E_h = \tau T \eta \sum_{i=1}^{N} p h_i^h S_i^h l(d_i), 
\end{equation}
where $p$ denotes the UAVs' transmit power, $\eta < 1$ is the efficiency of the RF-to-DC conversion and $l(\cdot), h_i^h, S_i^h$ are defined in subsection II.C. 

\subsubsection{Communication sub-slot}
In the transmission sub-slot, the IoT devices receive the information from their associated serving UAV-BS. Considering the nearest neighbor association policy, the receiver is served by the nearest UAV-BS, with $r_s=r_1$ denoting the serving distance. Once the communication link between the receiver and the serving BS has been established and since all UAV-BSs share the same resource blocks, the remaining active UAV-BSs are assumed to interfere the receiver{ This is rather common when no coordination among the UAV-BS is applied and all UAV-BSs try to communicate with an IoT device.}. In this case, the SINR at the receiver is given by 
\begin{equation}
SINR = \frac{ p h_i^c S_i^c l(r_s)}{ I  + \sigma^2}, 
\end{equation}  
where $I$ refers to the aggregate interference power and is given by $I = \sum_{i=1}^{N-1}  p h_i^c S_i^c l(v_i)$ where $v_i$ denotes the distance between the receiver and the $i$-th  interfering UAV-BSs and $\sigma^2$ is the additive white Gaussian noise power. 

\subsection{Performance Metrics}
\textbf{Definition 1} (Communication coverage probability $\mathcal{P}_{c}(\gamma_c)$). \emph{$\mathcal{P}_{c}(\gamma_c)$ is defined as the probability that the $SINR$ at the receiver, exceeds a predefined threshold $\gamma_c$, i.e., $\mathcal{P}_{c}(\gamma_c) \triangleq  \mathbb{P}(SINR > \gamma_c). $} 

\textbf{Definition 2} (Energy coverage probability  $\mathcal{P}_{h}(\gamma_h)$). \emph{ $\mathcal{P}_{h}(\gamma_h)$  is defined as the probability that the harvested energy $E_h$ at the receiver, exceeds a predefined threshold $\gamma_h$ required for circuit activation i.e., $\mathcal{P}_{h}(\gamma_h) \triangleq  \mathbb{P}(E_h > \gamma_h)$.} 

\textbf{Definition 3} (Joint energy and communication coverage probability). \emph{The receiver is said to be in joint coverage if i) $E_h \geq \gamma_h$, and ii) SINR $\geq \gamma_c$, i.e., $\mathcal{P}_{jc}(\gamma_h,\gamma_c) \triangleq  \mathbb{E}[\mathds{1}(E_h \geq \gamma_h) \mathds{1}({\rm{SINR}} \geq \gamma_c)]$, where $\mathds{1}(\cdot)$ denotes the indicator function.}

\section{Performance Analysis}

\subsection{Distance Distributions}
In this subsection, relevant distance distributions are derived as an intermediate step in the performance analysis.

\textbf{Lemma 1.} \textit{ The PDF of the serving distance $r_s=r_1$ is}
\begin{equation}
  f_{r_s}(r) = \frac{N!}{(N-1)!}\Big(1-\frac{\sqrt{r^2-h^2}}{R}\Big)^{N-1} \frac{r}{R \sqrt{r^2-h^2}}.
\end{equation}

%\begin{equation}
 % f_{r_s}(r) = \frac{N! \,  r (\sqrt{r^2-h^2})^{n-2} (R-\sqrt{r^2-h^2})^{N-n}}{R^N \, \Gamma(N-n+1) %\Gamma(n)} , \quad  h \leq r \leq \sqrt{h^2+R^2}.
%\end{equation}

\textit{Proof.} The full proof for deriving Lemma 1 follows from \cite{frontiers} and thus it is omitted here for brevity. \QEDA 

\textbf{Lemma 2.} \emph{ Conditioned on the serving distance $r_s = r$, the PDF of the unordered set of distances $\{v_i\}_{i=1:N-1}$ between the receiver and the $N-1$ UAV-BSs is given by }
\begin{equation}
  f_{{v_i}|r}(v_i) = \frac{v_i}{(R-\sqrt{r^2-h^2})(v_i^2-h^2)}, v_i \in [r,\sqrt{h^2+R^2}].
\end{equation}

\textit{Proof.}  Let $\{u_i\}_{i=1:N}$ denote the horizontal distance between the receiver and the projection of the $i$-th UAV on the ground. Moreover, the cumulative distribution function (CDF) of each element given by $F_{u_i}(u_i)=\frac{u_i}{R}$, $u_i \in [0,R]$. Then the CDF of the unordered set of distances $\{d_i\}_{i=1:N}$ are independently and identically distributed (i.i.d.) with CDF of each element given by  $F_{d_i}(d_i)=\frac{\sqrt{d_i^2-h^2}}{R}, \quad d_i \in [h,\sqrt{h^2+R^2}]$. The PDF  $f_{d_i}(d_i)$ can simply be obtained as  $f_{d_i}(d_i)=\frac{{\rm{d}}F_{d_i}(d_i)}{{\rm{d}}d_i}$, and is given by $f_{d_i}(d_i)= \frac{d_i}{R\sqrt{d_i^2-h^2}}$. Now $f_{{v_i}|r}(v_i)$ is given by $f_{{v_i}|r}(v_i)=\frac{f_{d_i}(v_i)}{1-F_{d_i}(r)}$ and after substituting $F_{d_i}(d_i)$ and $f_{d_i}(v_i)$, Lemma 2 yields. \QEDA

\subsection{Energy Coverage Probability}

In the following proposition, an expression of the energy coverage probability harvested from all UAV-BSs is derived in exact form. 

\textbf{Proposition 1.} \emph{The exact energy coverage probability harvested from all UAV-BSs is given by }
\begin{equation}
\mathcal{P}_{h}(\gamma_h) = 1-\Big[\mathcal{L}^{-1}\Big\{\frac{1}{s} \mathcal{L}_{E_h}(s)\Big\}(t)\Big]_{t=\gamma_{h}}, 
\end{equation}
\emph{ where $\mathcal{L}_{E_h}(s)$ is given by }
\begin{equation}
\begin{split}
\mathcal{L}_{E_h}(s) &= \Big[ \int_{0}^{\infty} \int_{h}^{\sqrt{h^2+R^2}} \Big( 1 + \frac{s\, p\, \tau \eta T S_i^h\, l(d_i)}{m}\Big)^{-m} f_{d_i}(d_i) f_{S_i^h}(S_i^h) {\rm{d}} d_i  {\rm{d}} S_i^h  \Big]^{N}.    
\end{split}  
\end{equation}

\textit{Proof.}  The proof builds upon the proof presented in \cite{frontiers} and thus it is omitted here for brevity. \QEDA

Unfortunately, the joint coverage probability cannot be obtained in exact form by exploiting Proposition 1. To obtain $\mathcal{P}_{jc}(\gamma_h,\gamma_c)$, the conditional energy coverage probability is first derived in the following subsection using the approximation of the Gamma distribution. The shape and scale parameters of the gamma distribution are estimated using the moment matching (MoM) technique\footnote{Analytical results were also obtained  using the maximum likelihood estimation (MLE) technique.It was observed that MLE results in a slightly tighter approximation on the energy coverage probability at a cost of extremely higher analytical complexity.}. 

\textbf{Lemma 3.} \emph{Conditioned on $r_s = r$, the conditional harvested energy $E_{h}|_{r} =  \sum_{i=2}^{N} p \eta \tau T S_i^h h_i^h l(v_i)$ from the rest UAV-BSs  is given by}
\begin{equation}
E_{h}|_{r} \, \, \approxequiv \, \, \Gamma(k_{mom}|_{r},\theta_{mom}|_{r}), 
\end{equation}
\emph{where  $k_{mom}|_{r}$ is given by $k_{mom}|_{r} = \frac{(\bar{E}(r))^2}{{\rm{var}}[E_{h}|_{r}]}$ and   $\theta_{mom}|_{r} = \frac{{\rm{var}}[E_{h}|_{r}]}{\bar{E}(r)}$, where ${\rm{var}}[E_{h}|_{r}] =  \mathbb{E}[(E_{h}|_{r})^2 | r] - (\bar{E}(r))^2$ and $\mathbb{E}[(E_{h}|_{r})^2 | r],$ are respectively given by}
\begin{equation} 
\begin{split}
&\mathbb{E}[(E_{h}|_{r})^2 | r] =  \underbrace{\int_{0}^{\infty} \dots \int_{0}^{\infty}}_{2N-2} \underbrace{\int_{r}^{\sqrt{h^2+R^2}} \dots \int_{r}^{\sqrt{h^2+R^2}}}_{N-1} \sum_{k_2,...,k_N \atop k_2+...+k_N=2}^{N} \binom{2}{k_2,...,k_N} \prod_{i=2}^{N} ( p \eta \tau T S_i^h h_i^h l(v_i))^{k_i} \\
&\times f_{{v_i}|r}(v_i)  f_{S_i^h}(S_i^h) f_{h_i^h}(h_i^h) {\rm{d}} v_i  {\rm{d}} S_i^h {\rm{d}} h_i^h, \\
&\bar{E}(r) = (N-1) p \eta \tau T \int_{0}^{\infty} \int_{r}^{\sqrt{h^2+R^2}} l(v_i) S_i^h f_{S_i^h}(S_i^h) f_{{v_i}|r}(v_i) {\rm{d}} v_i  {\rm{d}} S_i^h,
\end{split}
\end{equation}
\emph{where $\binom{2}{k_2,...,k_N} = \frac{2!}{k_2!...k_N!}$ is the multinomial coefficient.}

\emph{Proof.} See Appendix A. \QEDA 

\subsubsection{Conditional Energy Coverage Probability}

To analyze the energy harvested by the receiver, the conditional energy coverage probability of the receiver can now be derived in the following Proposition.  

\textbf{Proposition 2.} \emph{ Conditioned on $r_s = r$, the conditional energy coverage probability $\mathcal{P}_{h }(\gamma_h|r)$ is given by }
\begin{equation}
\begin{split}
\mathcal{P}_{h }(\gamma_h|r) &=  \int_{0}^{\infty} \int_{0}^{\infty}  \frac{\Gamma\big(k_{mom}|_{r},\frac{[\gamma_h - p \eta \tau T S_i^h h_i^h  l(r)]^{+}}{\theta_{mom}|_{r}}\big)}{k_{mom}|_{r}} f_{S_i^h}(S_j^h) f_{h_i^h}(h_i^h) {\rm{d}} h_i^h {\rm{d}} S_i^h,
\end{split}
\end{equation}
\emph{where $\Gamma(\cdot,\cdot)$ denotes the upper incomplete Gamma function and} 
$[x]^{+} = \max\{x,0\}$.
  
\emph{Proof.} See Appendix B.  \QEDA

%\emph{ where $\bar{E}(r)$ is given by }
%\begin{equation}
% \bar{E}(r) = (N-1) \int_{0}^{\infty} \int_{r}^{\sqrt{h^2+R^2}} p_{on}\, x_i\, (v_i|r)^{-\alpha} f_{{v_i}|r}(v_i)  f_{S_i^h}(x_i) {\rm{d}} v_i  {\rm{d}} x_i, 
%\end{equation}
%\emph{where} $C(\gamma_h) = \frac{\gamma_h}{p\,T\tau K \eta}$ 

%\textbf{Proposition 2.} \emph{ Conditioned on $r_s = r$, the conditional energy coverage probability $\mathcal{P}_{h }(\gamma_h|r)$ is given by }
%\begin{equation}
%\mathcal{P}_{h }(\gamma_h|r) = \int_{0}^{\infty} \sum_{k=0}^{m-1} \frac{(m r^{\alpha} [C(\gamma_h)-\bar{E}(r)]^{+})^k}{k!} e^{- m r^{\alpha} [C(\gamma_h)-\zeta(r)]^{+} } {\rm{d}} x_i, 
%\end{equation}
%\emph{ where $\bar{E}(r)$ is given by }
%\begin{equation}
% \bar{E}(r) = (N-1) \int_{0}^{\infty} \int_{r}^{\sqrt{h^2+R^2}} p_{on}\, x_i\, (v_i|r)^{-\alpha} f_{{v_i}|r}(v_i)  f_{S_i^h}(x_i) {\rm{d}} v_i  {\rm{d}} x_i, 
%\end{equation}
%\emph{where} $C(\gamma_h) = \frac{\gamma_h}{p\,T\tau K \eta}$ \emph{and},$ [x]^{+} = \max\{x,0\}$.

%\emph{Proof.}  See Appendix C. \QEDA 

\subsection{Communication Coverage Probability}
The Laplace transform of the aggregate interference power distribution is now obtained as intermediate step in the communication coverage probability analysis. 

\textbf{Lemma 4.} \emph{The Laplace transform of the aggregate interference power distribution conditioned on the serving distance $r_s=r$ is given by}
\begin{equation} 
\begin{split}
\mathcal{L}_{I}(s|r)&=  \Big[  \int_{0}^{\infty} \int_{r}^{\sqrt{h^2+R^2}}  \Big( 1 + \frac{s\, p\,  S_i^c\, {v_i^{-\alpha}}}{m_I}\Big)^{-m_I} f_{{v_i}|r}(v_i)  f_{S_i^c}(S_i^c) \,{\rm{d}}v_i {\rm{d}}S_i^c\Big]^{N-1}.
\end{split}
\end{equation}

\textit{Proof.} The proof follows similar conceptual lines as the one presented in \cite{globecom} and thus it is omitted here for brevity. \QEDA

\textbf{Proposition 3.} \emph{Conditioned on $r_s=r$, the communication coverage probability $\mathcal{P}_{c}(\gamma_c| r)$ is given by} 
\begin{equation} 
\begin{split}
\mathcal{P}_{c}(\gamma_c| r) &=\int_{0}^{\infty} \sum_{k=0}^{m - 1}  \frac{(-1)^k}{k!} \Big(\frac{{m \gamma }}{p l(r) S_i^c}\Big)^k  {\Bigg[ \frac{\partial^{k} \mathcal{L}_{I_{tot}}(s|r)}{\partial{s^k}}\Bigg]}_{s=\frac{m \gamma}{p l(r)  S_i^c}}  f_{S_i^c}(S_i^c)    {\rm{d}} S_i^c ,
\end{split}
\end{equation}
\emph{where} $\mathcal{L}_{I_{tot}}(s|r) = \exp({-\sigma^2 s)}  \mathcal{L}_I(s|r)$.  

\textit{Proof.} The conditional coverage probability can be obtained  as
\begin{equation} 
\begin{split}
&\mathcal{P}_{c}(\gamma_c |r ) \\
&= \mathbb{P}\Big[ \frac{ p h_i^c S_i^c l(r)}{ I  + \sigma^2} > \gamma_c |r ]\\
& = \mathbb{P}\Big[   h_i^c > \frac{\gamma_c (I  + \sigma^2)}{p   l(r)} S_i^c |r ] \\
& \myeqa \mathbb{E}_{S_i^c,I}\Big[\sum_{k=0}^{m - 1} \frac{1}{k!} \Big(-\frac{{m \gamma_  (I  + \sigma^2)c}}{p l(r) S_i^c}\Big)^k e^{-\frac{{m \gamma_c  (I  + \sigma^2)}}{p l(r) S_i^c}}\Big]\\
& \myeqb \mathbb{E}_{S_i^c}\bigg[\sum_{k=0}^{m - 1} \frac{(-1)^k}{k!}  \Bigg(\frac{{m \gamma_c}}{p l(r) S_i^c}\Bigg)^k  {\Bigg[ \frac{\partial^{k} \mathcal{L}_{I_{tot}}(s|r)}{\partial{s^k}}\Bigg]}_{s=\frac{m \gamma_c }{p l(r) S_i^c}}\Bigg], 
\end{split}
\end{equation}
where (a) follows from writing the cCDF of the gamma random variable $h_i^c$ as a sum for integer values of $m$, and (b) follows after applying the formula $ \mathbb{E}_{X}[x^k \exp(-s X)] = (-1)^k \frac{\partial^{k} \mathcal{L}_{X}(s)}{\partial{s^k}}$. Finally, (14) yields immediately after applying the expectation over the random variable $S_i^c$.  \QEDA

\subsection{Joint Energy and Communication Coverage Probability}
Conditioned on $r_s=r$, the conditional energy and communication coverage probability have been derived. Therefore, the joint energy and communication coverage probability can now be derived in the following Theorem. 

\textbf{Theorem 1.} \emph{The joint energy and communication coverage probability $\mathcal{P}_{jc}(\gamma_h,\gamma_c)$ of a receiver in a UAV corridor-assisted IoT network is given by}
\begin{equation} 
\begin{split}
&\mathcal{P}_{jc}(\gamma_h,\gamma_c) =  \int_{h}^{\sqrt{h^2+R^2}} \mathcal{P}_{h }(\gamma_h|r) \mathcal{P}_{c}(\gamma_c|r) f_{r_s}(r) {\rm{d}} r.
\end{split}
\end{equation}

\textit{Proof.} Recalling Definition 3, the joint energy and communication coverage probability is given by
\begin{equation} 
\begin{split}
&\mathcal{P}_{jc}(\gamma_h,\gamma_c) = \mathbb{E}[\mathds{1}(E_h \geq \gamma_h) \mathds{1}({\rm{SINR}} \geq \gamma_c)] \\
&\myeqa \mathbb{E}_{r_s}[\mathbb{E}_{h_i^h,S_i^h}[\mathds{1}(E_h \geq \gamma_h) | r] \times \mathbb{E}_{h_i^c,S_i^c}[ \mathds{1}({\rm{SINR}} \geq \gamma_c) |r]] \\
&  \myeqb  \mathbb{E}_{r_s}[\mathbb{P}(E_h\geq \gamma_h |r) \times \mathbb{P}({\rm{SINR}}\geq \gamma_c |r)],
\end{split}
\end{equation}
where (a) follows from independence between the fading channels, shadowing conditions as well as the state of the UAV-BSs during the two sub-slots.  Next, (b) follows directly after applying the expectations which result to the conditional probabilities that have been proved in Proposition 2 and Proposition 3. After substituting Proposition 2 and Proposition 3 in (17) and applying the expectation over $r_s$ w.r.t $f_{r_s}(r)$ given by Lemma 1, Theorem 1 yields. \QEDA

 \begin{figure}[!t]
    \centering
    \includegraphics[keepaspectratio,width=3.0in]{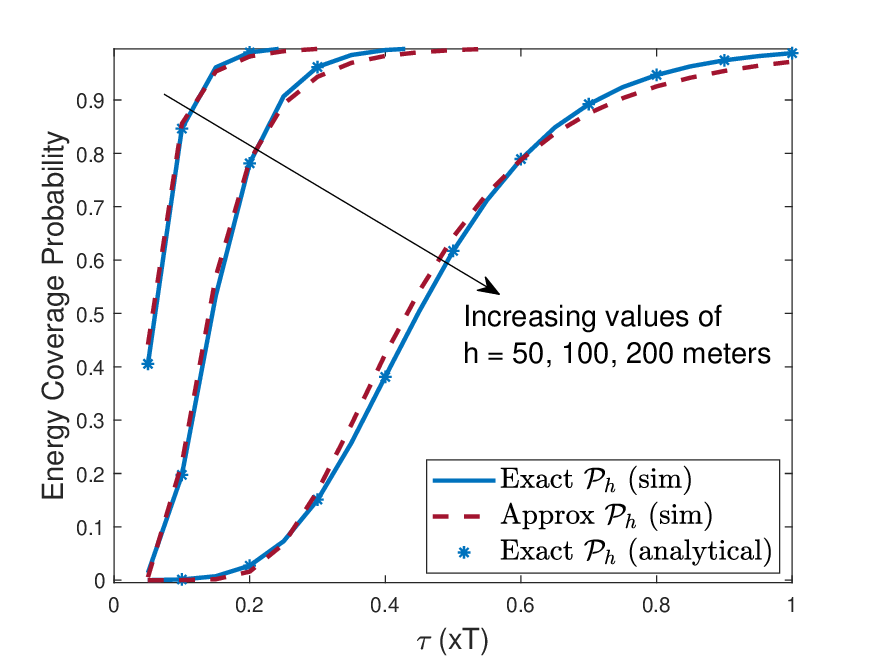}
       \caption{ Energy coverage probability versus $\tau$ for different values of $h$.}
    \label{fig:Fig2}
 \includegraphics[keepaspectratio,width=3.0in]{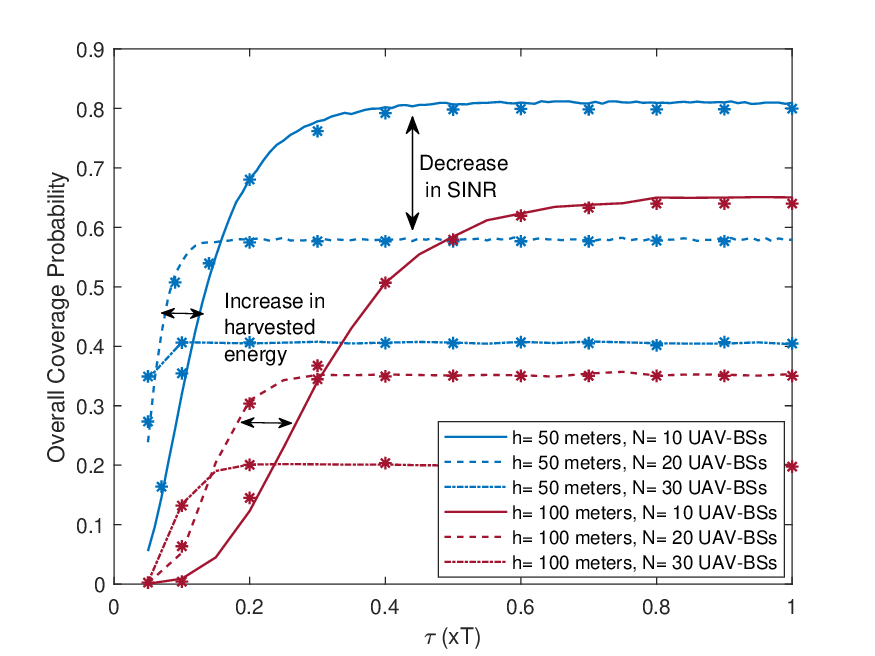}
       \caption{ Overall coverage probability versus $\tau$ for different values of $h$ and $N$.}
    \label{fig:Fig3}
\end{figure}

\section{Results and Discussion}
In this section, numerical results are presented to evaluate the performance achieved in a UAV corridor-assisted RF-powered IoT network in terms of energy coverage probability, communication coverage probability and overall coverage probability. The accuracy of the analytical results is verified by comparing them with the empirical results obtained from Monte-Carlo simulations. Unless otherwise stated, the default system parameters are as follows: $N= 10$, $h=100$ meters, $R= 200$ meters, $\alpha=2.2$, $f_c = 3.5$GHz, $m = 2$, $q = 3$, $p = 32$ dBm, $\eta=0.5$, $T = 1$ sec and $\tau = T/4$.  

%\begin{figure}[!tbp]
 % \centering
 % \begin{minipage}[b]{0.45\columnwidth}
 %    \includegraphics[trim=0 0 0 0,clip,width=\textwidth]{Eh_vs_tau.eps}
  %  \caption{ Energy coverage probability versus $\tau$ for different number of $h$.}
 % \end{minipage}
%  \hfill
%  \begin{minipage}[b]{0.45\columnwidth}
 %    \includegraphics[trim=0 0 0 0,clip,width=\textwidth]{jointcov_vs_tau.eps}
  %  \caption{ Overall coverage probability versus $\tau$,  for different values of $h$ and $N$. (Markers denote the analytical results.)}
 % \end{minipage}
%\end{figure}

Fig. 2 presents the exact energy coverage probability $\mathcal{P}_{h}$, obtained by Proposition 1, and the approximated one, obtained by deconditioning over $r$ in Lemma 3, versus $\tau$ for different values of $h$. A first observation is that $\mathcal{P}_{h}$ decreases with the increase of $h$ due to the higher path-losses. Moreover, as the time duration for charging purposes increases, higher probability is observed that the receiver reaches its energy threshold. Finally, it is observed that Gamma approximation provides an extremely tight approximation of the exact expression for $\mathcal{P}_{h}$. 

Fig. 3 depicts the overall coverage probability $\mathcal{P}_{jc}$, obtained by Theorem 1, versus $\tau$, for different values of $h$ and $N$. For a given value of $h$, it is observed that by increasing the number of deployed UAV-BSs, the overall coverage probability decreases due to the higher interference experienced during the communication stage. Nevertheless, the IoT devices reach their energy target threshold faster until an upper bound is reached, which is now determined by the communication performance. Moreover, the enhancement in energy coverage dominates over the reduction in communication coverage probability. When $\tau$ reaches a certain value, for which the IoT devices are fully charged, i.e., $\mathcal{P}_{h} =1$, the communication coverage probability determines the overall coverage performance. In such cases, interference management and/or coordination schemes are of pivotal sense.

%\begin{figure}[!tbp]
 % \centering
 % \begin{minipage}[b]{0.45\columnwidth}
 %    \includegraphics[trim=0 0 0 0,clip,width=\textwidth]{jointcov_vs_N.eps}
  %  \caption{Overall coverage probability versus $N$ for different values of $\tau$. (Markers denote the analytical results.) }
 % \end{minipage}
 % \hfill
  %\begin{minipage}[b]{0.45\columnwidth}
 %    \includegraphics[trim=0 0 0 0,clip,width=\textwidth]{heatmap.eps}
  %  \caption{Heatmap of overall coverage probability versus $R$ and $h$ for different values of $N$. %(Markers denote the optimal sets $(R,h)$ for maximizing the overall coverage performance.)}
 % \end{minipage}
%\end{figure}

%\begin{figure}[h!]
%\begin{center}
%\includegraphics[width=10cm]{jointcov_vs_N.eps}% This is a *.eps file
%\end{center}
%\caption{ Overall coverage probability versus $N$ for different values of $\tau$. (Markers denote %the analytical results.) }\label{fig:1}
%\end{figure}

%\begin{figure}[h!]
%\begin{center}
%\includegraphics[width=10cm]{heatmap.eps}% This is a *.eps file
%\end{center}
%\caption{ Heatmap of overall coverage probability versus $R$ and $h$ for different values of $N$. (Markers denote the optimal sets $(R,h)$ for maximizing the overall coverage performance.) }\label{fig:1}
%\end{figure}

\begin{figure}[!t]
\centering
\includegraphics[keepaspectratio,width=3.0in]{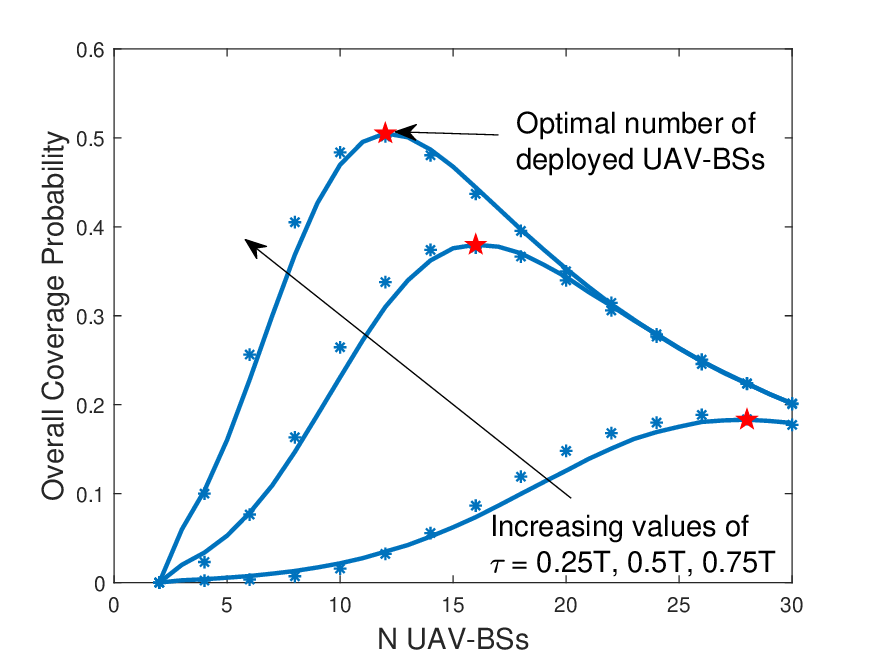}
\caption{  Overall coverage probability versus $N$ for different values of $\tau$. Red stars denote the optimal number of deployed UAV-BSs for maximizing the joint coverage performance. (Markers denote the analytical results.)}
\label{fig:Fig4}
\end{figure}

\begin{figure}[!t]
\centering
\captionsetup{font=small} % Set the font size for the main caption
\subfloat[\scriptsize{$N=10$ \rm{UAV-BSs}}]{\includegraphics[width=2.5in]{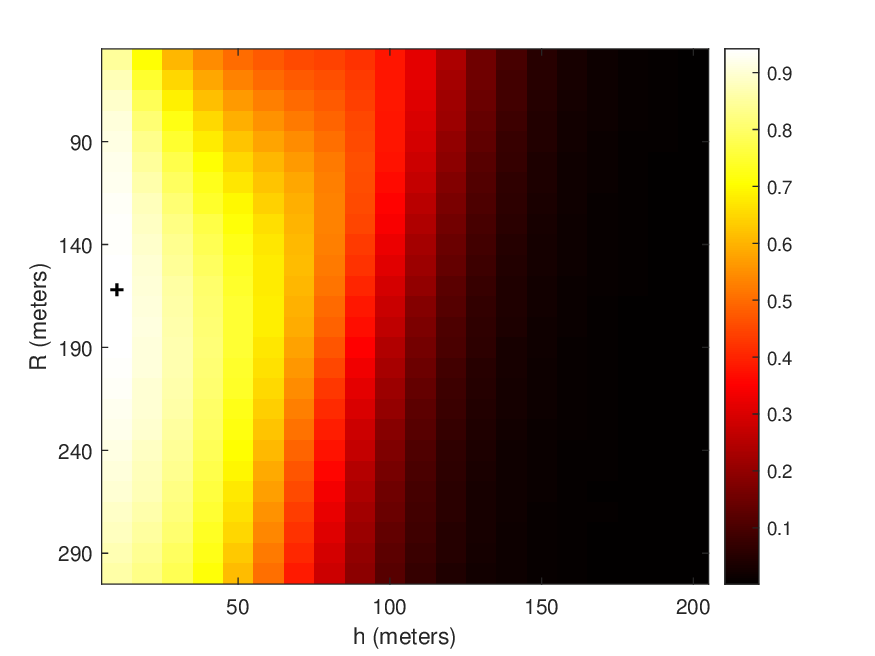} 
\label{fig_first_case}}\\
\subfloat[\scriptsize{$N=20$ \rm{UAV-BSs}}]{\includegraphics[width=2.5in]{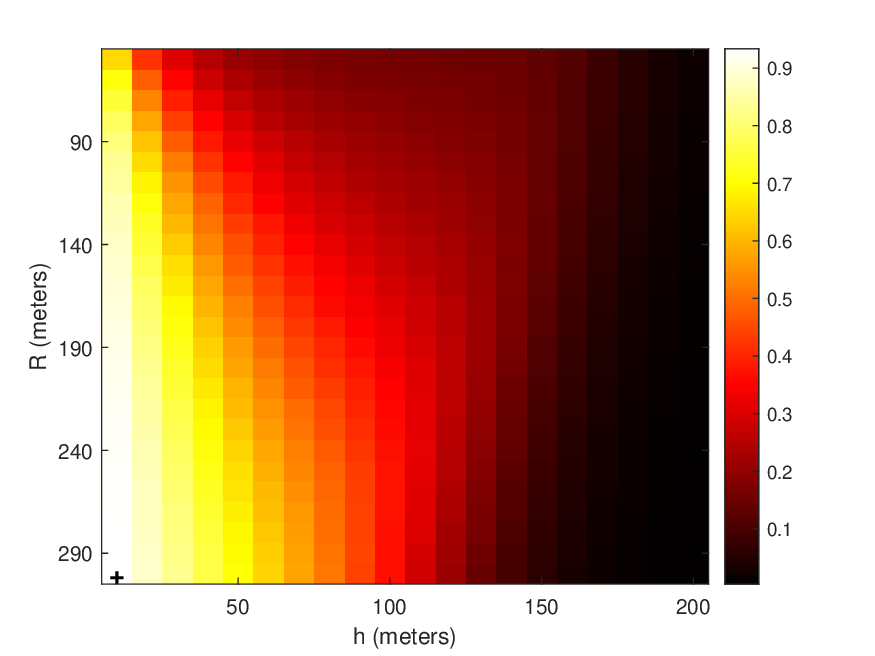}%
\label{fig_second_case}}
\caption{Heatmap of overall coverage probability versus $R$ and $h$ for different values of $N$. (Markers denote the optimal sets $(R,h)$ for maximizing the overall coverage performance.)}
\label{fig_sim}
\end{figure}

% \begin{figure}[!t]
   % \centering
   % \includegraphics[keepaspectratio,width=3in]{jointcov_vs_N_demandingEh_v2.eps}
    %   \caption{  Overall coverage probability versus $N$ for different values of $\tau$. Red stars denote the optimal number of deployed UAV-BSs for maximizing the joint coverage performance. (Markers denote the analytical results.)}
  %  \label{fig:Fig4}
%\vspace{0.1cm}
  %  \includegraphics[keepaspectratio,width=2.8in]{heatmapv2.eps}
    %   \caption{ Heatmap of overall coverage probability versus $R$ and $h$ for different values of $N$. (Markers denote the optimal sets $(R,h)$ for maximizing the overall coverage performance.) }
  %  \label{fig:Fig5}
%\end{figure}

%\begin{figure}[!tbp]
 % \centering
%  \begin{minipage}[b]{0.45\columnwidth}
 %    \includegraphics[trim=0 0 0 0,clip,width=\textwidth]{jointcov_vs_N.eps}
 %   \caption{ Overall coverage probability versus $N$ for different values of $\tau$. Red stars denote the optimal number of deployed UAV-BSs for maximum joint coverage performance. (Markers denote the analytical results.)}
 % \end{minipage}
 % \hfill
 % \begin{minipage}[b]{0.45\columnwidth}
  %   \includegraphics[trim=0 0 0 0,clip,width=\textwidth]{heatmap.eps}
  %  \caption{ Heatmap of overall coverage probability versus $R$ and $h$ for different values of $N$. (Markers denote the optimal sets $(R,h)$ for maximizing the overall coverage performance.) }
%  \end{minipage}
%\end{figure}

Fig. 4 presents the overall coverage probability $\mathcal{P}_{jc}$ versus the number of deployed UAV-BSs $N$, for different values of $\tau$. It is observed that the more time is devoted to energy harvesting, less number of UAV-BSs are required for optimizing the overall coverage performance. This is attributed to the fact that although an increased number of deployed UAV-BSs fully charges the IoT devices, it also makes the interference power at the receiver severer. In this case, there is an optimal value $N^{*}$ for $N$, shown by a red star in Fig.4, which ensures the best balance in the trade-off between $\mathcal{P}_{h}$  and $\mathcal{P}_{c}$. When $N > N^{*}$, the overall IoT network's performance decreases. Indeed, once the IoT devices are fully charged, an increase in $N$ just reduces $\mathcal{P}_{c}$ due to severer interference. 

Fig. 5 presents the heatmap of the overall coverage probability $\mathcal{P}_{jc}$ versus $R$ and $h$, for different values of $N$. The optimal sets of $(R,h)$ for maximizing the overall coverage probability are also depicted with black markers. Interestingly, as $N$ increases the optimal UAV deployment radius $R$ is shifted towards the higher radius. This is because with the increase of $N$, i) the probability that the IoT devices are fully charged increases despite the larger values of $R$, and ii) the interference power experienced during the communication phase deteriorates.

%\begin{figure*}[!htbp] 
%\centering
%\begin{subfigure}{.5\columnwidth}
%\includegraphics[width=\columnwidth]{Eh_vs_tau.eps}%
%\caption{$r \leqslant R_{hex} $}%
%\label{subfiga}%
%\end{subfigure}
%\hfill%
%\begin{subfigure}{.5\columnwidth}
%\includegraphics[width=\columnwidth]{jointcov_vs_tau.eps}%
%\caption{$R_{hex} < r \leqslant R_{sq}$}%
%\label{subfigb}%
%\end{subfigure}
%\hfill%
%\begin{subfigure}{.5\columnwidth}
%\includegraphics[width=\columnwidth]{jointcov_vs_N.eps}%
%\caption{$R_{sq} < r \leqslant R_{mid}$}%
%\label{subfigc}%
%\end{subfigure}%
%\hfill
% \begin{subfigure}{.5\columnwidth}
%\includegraphics[width=\columnwidth]{heatmap.eps}%
%\caption{$r = R_{mid}$}%
%\label{subfigd}%
%\end{subfigure}% 
%\end{figure*} 

\section{Conclusions}
This letter investigated the joint energy and SINR-based coverage probability in UAV corridor-assisted RF-powered IoT networks in the presence of shadowing. The locations of UAV-BSs in the aerial corridor were modeled as a 1D BPP. Our results showed that devoting more time for energy harvesting can reduce the need for dense deployment of UAV-BSs required to fully charge the IoT devices, thereby resulting in a balanced trade-off between energy and SINR coverage performance. Moreover, it was  shown that as the number of deployed UAV-BSs increases, the optimal UAV deployment radius $R$ that maximizes the overall coverage performance increases as well. A direct extension is to extend the framework so as to include multiple UAV \emph{lanes} in the 2D plane.

\appendices
\section{Proof of Lemma 3}

To obtain $\bar{E}(r)$, we initially ignore the sum in $E_{h}|_{r}$ and we assume only one term, i.e., $E^i_{h}|_{r} = p \eta \tau T S_i^h h_i^h l(v_i)$. Therefore, $\mathbb{E}[E^i_{h}|_{r}] $  is first obtained as
\begin{equation} 
\begin{split}
&\mathbb{E}[E^i_{h}|_{r}] =  \mathbb{E}[ p \eta \tau T S_i^h h_i^h l(v_i)] = p \eta \tau T \mathbb{E}_{S_i^h, h_i^h, l(v_i) }[ S_i^h h_i^h l(v_i)] \\
&\myeqa p \eta \tau T \mathbb{E}_{ h_i^h, l(v_i) }\Big[ \int_{0}^{\infty} S_i^h  f_{S_i^h}(S_i^h) h_i^h l(v_i) {\rm{d}} S_i^h \Big] \\
&\myeqb p \eta \tau T \mathbb{E}_{ l(v_i) }\Big[ \int_{0}^{\infty} S_i^h  f_{S_i^h}(S_i^h) l(v_i) {\rm{d}} S_i^h \Big] \\
&\myeqc p \eta \tau T \int_{r}^{\sqrt{h^2+R^2}} \int_{0}^{\infty} S_i^h  f_{S_i^h}(S_i^h) l(v_i) f_{{v_i}|r}(v_i) {\rm{d}} S_i^h {\rm{d}} v_i,
\end{split}
\end{equation}
where (a) follows independence between the random variables $S_i^h, h_i^h, l(v_i)$ and after applying the definition of the expected value in $S_i^h$, (b) follows from the fact that $\mathbb{E}[h_i^j]=m \frac{1}{m}=1$, and (c) follows after applying the definition of the expected value in $l(v_i)$. Finally, notice that $E_{h}|_{r}$ is a sum of $N-1$  i.i.d. random variables $E^i_{h}|_{r}$, i.e.,  $E_{h}|_{r} =  \sum_{i=2}^{N} E^i_{h}|_{r}$. Then, $\bar{E}(r) = (N-1) \mathbb{E}[E^i_{h}|_{r}]$. Next, to obtain $\mathbb{E}[(E_{h}|_{r})^2 | r]$, we first derive $(E_{h}|_{r})^2$ in analytically tractable form. Therefore, $(E_{h}|_{r})^2$ can be rewritten as
\begin{equation} 
\begin{split}
&(E_{h}|_{r})^2 = \Big(\sum_{i=2}^{N} p \eta \tau T S_i^h h_i^h l(v_i)\Big)^2 \\
&= (p \eta \tau T S_2^h h_2^h l(v_2)+ p \eta \tau T S_3^h h_3^h l(v_3) +...+ p \eta \tau T S_N^h h_N^h l(v_N))^2\\
& \myeqa \sum_{k_2,...,k_N \atop k_2+...+k_N=2}^{N} \binom{2}{k_2,...,k_N} \prod_{i=2}^{N} ( p \eta \tau T S_i^h h_i^h l(v_i))^{k_i}, 
\end{split}
\end{equation}
where (a) follows from the multinomial theorem, i.e., by exploiting the formula $(v_1+...+v_m)^n = \sum_{k_1,...,k_m \atop k_1+...+k_m=n}^{n}  \binom{n}{k_1,...,k_m} v_1^{k_1} v_2^{k_2}...v_m^{k_m}$. Having derived $(E_{h}|_{r})^2$ in a compact form, $\mathbb{E}[(E_{h}|_{r})^2 | r]$ yields immediately after following similar conceptual lines for deriving (18) and applying the definition of expectation for the product of i.i.d. random variables. Finally, having derived $\mathbb{E}[E^i_{h}|_{r}]$ and $\mathbb{E}[(E_{h}|_{r})^2 | r]$, the parameters $k_{mom}|_{r}$ and $\theta_{mom}|_{r}$ are obtained as shown in (10), which completes the proof. 

\section{Proof of Proposition 2}
Conditioned on the energy harvested from the serving BS at distance $r_s=r$, the conditional energy coverage probability is given by
\begin{equation}
\begin{split}
&\mathcal{P}_{h}(\gamma_h|r) = \mathbb{P}[E_h>\gamma_h|r] \\
&\myeqa  \mathbb{P}\Big[ p \eta \tau T S_i^h h_i^h l(r)+ \sum_{i=2}^{N} p \eta \tau T S_i^h h_i^h l(v_i) > \gamma_h \Big] \\
& \myeqb  \mathbb{P}[ p \eta \tau T S_i^h h_i^h l(r) + \Gamma(k_{mom}|_{r},\theta_{mom}|_{r}) > \gamma_h ] = \mathbb{P}[ \Gamma(k_{mom}|_{r},\theta_{mom}|_{r}) > \gamma_h - p \eta \tau T S_i^h h_i^h l(r)] \\
&\myeqc \mathbb{P}[ \Gamma(k_{mom}|_{r},\theta_{mom}|_{r}) > [\gamma_h - p \eta \tau T S_i^h h_i^h l(r)]^+]\\
&= 1- \mathbb{P}[ \Gamma(k_{mom}|_{r},\theta_{mom}|_{r}) < [\gamma_h - p \eta \tau T S_i^h h_i^h l(r)]^+]\\
&\myeqd \mathbb{E}_{S_i^h, h_i^h}[\Gamma\big(k_{mom}|_{r},\frac{[\gamma_h - p \eta \tau T S_i^h h_i^h  l(r)]^{+}}{\theta_{mom}|_{r}}\big)]\\
&\myeqe \int_{0}^{\infty} \int_{0}^{\infty}  \frac{\Gamma\big(k_{mom}|_{r},\frac{[\gamma_h - p \eta \tau T S_i^h h_i^h  l(r)]^{+}}{\theta_{mom}|_{r}}\big)}{k_{mom}|_{r}} f_{S_i^h}(S_j^h) f_{h_i^h}(h_i^h) {\rm{d}} h_i^h {\rm{d}} S_i^h,
\end{split}
\end{equation}
where (a) follows from independence between the distances $v_i$ and the random variables $S_i^h$ and $h_i^h$, (b) follows directly from Lemma 3, (c) follows from the definition of the Gamma random and make sure that the argument is always positive, (d) follows after exploiting the formula $\frac{\Gamma(m,m x)}{\Gamma(m)} = 1-\frac{\gamma(m,m x)}{\Gamma(m)}$ in the complementary CDF (cCDF) of the Gamma random variable, where $\gamma(\cdot)$ denotes the generalized incomplete gamma function. Subsequently, we apply in (d) the definition of the probability w.r.t. the random variables $S_i^h$ and $h_i^h$, and (e) follows from applying the definition of the expectation and averaging over $S_i^h$ and $h_i^h$ w.r.t. their PDFs $ f_{S_i^h}(S_j^h)$ and $f_{h_i^h}(h_i^h)$, respectively.

 \bibliography{IEEEabrv,references}

\end{document}